\documentclass[11pt,a4paper]{article}  
\usepackage[english]{babel} 
\usepackage{graphicx}
\usepackage{epstopdf}
\usepackage{amstext,amsfonts,amsbsy,amssymb,bbm}

\newcommand {\be}[1]{\begin{eqnarray} \mbox{$\label{#1}$}  }
\newcommand{\ee}{\end{eqnarray}}

\newcommand{\pref}[1]{(\ref{#1})}

\newcommand\ie {{\it i.e.}, }
\newcommand\eg {{\it e.g. }}

\newcommand{\nn}{\nonumber\\}

\newcommand\ra{\rightarrow }

\newcommand\half{\frac 1 2 }

\newcommand{\mean}[1]{\left \langle #1 \right \rangle}
\renewcommand{\v}[1]{{\bf #1}}
\newcommand{\ket}[1]{|#1\rangle}
\newcommand{\bra}[1]{\langle #1 |}
\newcommand{\braket}[2]{\langle #1|#2 \rangle}

\newcommand{\dd}[2]{{d{#1}\over d{#2}}}

\newcommand{\cA}{ {\cal A} }
\newcommand{\cO}{ {\cal O} }

\newcommand{\ga}{ {\alpha} }
\newcommand{\gz}{ {\zeta} }
\renewcommand{\gg}{\gamma}

\begin{document}

\title{Explicit mapping between a 2D quantum Hall system and a 1D Luttinger liquid\\
\it{II. Correlation functions} }

\author{Mats Horsdal and Jon Magne Leinaas\\
Department of Physics,University of Oslo,\\ P.O.
Box 1048 Blindern, 0316 Oslo, Norway}

\date{March 19, 2007}

\maketitle
\begin{abstract} 
We study a simple model of a quantum Hall system with the electrons confined to a linear, narrow channel. The system is mapped to a 1D system which in the low-energy approximation has the form of a Luttinger liquid with different interactions between particles of equal and of opposite chiralities. In a previous paper (Part I) we studied this mapping at the microscopic level, and discussed the relation between the parameters of the 2D system and the corresponding 1D Luttinger liquid parameters. We follow up this study here and derive the low-energy form of the 2D electron correlation function by applying the mapping from the 1D correlation function. We examine in particular the modification of the asymptotic behaviour of the correlation function due to interactions and find for the 2D function a similar fall off with distance as in 1D.

\end{abstract}


\section{Introduction}

In Part I \cite{Horsdal07} we have examined in detail the relation between the quantum Hall system and the Luttinger liquid in a simplified case, where an explicit microscopic mapping between the two systems can be performed.  This is a case of integer filling where the 2D electron gas is constrained to a narrow channel by a harmonic oscillator potential, and where the electron interaction is  assumed to be sufficiently long range to interconnect the two edges.  The corresponding 1D system is in the low energy approximation described by a full Luttinger liquid  with both chiralities present, and with different interactions between particles of the same chirality and with opposite chiralities. In \cite{Horsdal07} we in particular focussed on how the Luttinger liquid parameters were modified by the electron interaction.

In the present paper we study the electron correlation function and show that an explicit form of the 2D correlation function can be found in the low energy approximation. As is well known the electron correlation function of the 1D Luttinger liquid can be derived by the bosonization technique \cite{LutherPeschel74,Mattis74,Haldane81} and we follow \cite{Haldane81} in order to establish this for the present case. 
The mapping discussed in \cite{Horsdal07} is then used to derive the 2D correlation function. The function is found to have a natural separation into two parts, which we associate with the bulk and the edge of the 2D electron gas.

We focus in particular on the effects of the electron interaction on the correlation function. A part of the motivation for this is the recent discussion about the effects of interactions on the electron correlation function in the fractional quantum Hall effects, where numerical studies \cite{Goldman01,Jain01,WanEvers04} have indicated, in some cases,  deviations in the asymptotic behaviour of the  calculated 2D correlation function from the expected behaviour based on the correspondence with the 1D chiral Luttinger liquid \cite{Wen92}. 
These studies in turn have been motivated by the difference between the observed tunneling resistance and the expected resistance based on the chiral Luttinger description of the quantum Hall edge \cite{Grayson98,Hilke01}. The discussion illustrates the point that the analysis of the edge effects to some extent has to rely on qualitative arguments  and on numerical studies of systems with a limited number of particles, and that this may leave room for some uncertainties due to different interpretations. 

The simplified model studied in this paper cannot be directly related to the unsettled questions concerning the edge effects at fractional filling. Even so, it seems interesting to examine in detail a case where an explicit mapping between the 2D quantum Hall system and a 1D Luttinger liquid can be performed, and to examine in this case how the electron correlation function of the quantum Hall system is related to the correlation function of the Luttinger liquid. As a particular result we find no difference between the asymptotic parameters in 1D and 2D, caused by the interactions. However, the region where deviation from the asymptotic  form is non-negligible, may extend to a distance that is rather large  when measured in units of magnetic length.

As discussed in \cite{Horsdal07} we make certain simplifying assumptions for the model studied. We assume a soft harmonic confining potential, with an oscillator frequency that is small compared to the cyclotron frequency, and also assume a weak particle interaction in the form of a gaussian potential.
In the following we first make a brief review of  this model and its mapping from 2D to 1D. We further discuss the electron correlation in the 1D description and show how the 2D correlation function can be derived. We supplement these results with some graphs that demonstrate how the electron correlation function and the density function in 2D is modified by the interactions for some specific choices of parameter values.

\section{The model}
The single particle part of the 2D electron Hamiltonian has the form
\be{Hamsp}
H&=&{1\over {2m}}(p_x+eBy)^2+{1\over {2m}}p_y^2+\half m\omega^2y^2
\ee
which describes spinless electrons in a constant perpendicular magnetic field $B$ and subject to a harmonic confining potential in the y-direction. Landau gauge has been chosen here, with vector potential
\be{Landaugauge}
A_x=-yB\;,\quad A_y=0
\ee
We re-write this in terms of the effective cyclotron frequency $\bar{\omega}_c=\sqrt{\omega_c^2+\omega^2}$, with $\omega_c=eB/m$, 
\be{Hamsp2}
H={1\over {2m}}p_y^2+\half m\bar{\omega}_c^2(y+\frac{\omega_c}{m\bar{\omega}_c^2}p_x)^2 +{1\over {2m}}\frac{\omega^2}{\bar{\omega}_c^2}p_x^2
\ee 
The condition for a soft confining potential, $\omega<<\omega_c$ is assumed in the following.

Since $p_x$ and $H$ commute they can simultaneously be quantized, and also the operator defined by the first two terms of \pref{Hamsp2} can be quantized simultaneous with $p_x$. This part of $H$ has the form of the Hamiltonian for an electron in the effective magnetic field $\bar B=\sqrt{B^2+m^2\omega^2/e^2}$. As a basic assumption we assume throughout the paper that  the low energy approximation is well satisfied, where electrons are confined to the lowest Landau level of the Hamiltonian for the effective magnetic field. This means that a complete set of eigenstates can be written in the form,
\be{pxeigen}
\psi_k(x,y)= {1\over{\sqrt{L}}} e^{i kx}\psi_0(y-y_k)
\ee
with  $\hbar k$ as the quantized value of $p_x$, $L$ as a normalization length in the x-direction and $y_k=-(\omega_c/\bar\omega_c)\bar l_B^2 k$ as a k-dependent shift in the y-direction. $\psi_0$ is the ground state wave function for the (effective) harmonic oscillator equation in the $y$-direction,
\be{harmosc}
\psi_0(y)=(\frac{1}{\pi \bar l_B^2})^{\frac{1}{4}}e^{-\frac{1}{2 \bar l_B^2}y^2}
\ee 
In the expressions above the magnetic length of the effective magnetic field has been introduced, $\bar l_B=\sqrt{\hbar /e\bar B}$

The restriction to the lowest Landau level effectively reduces the dimension of the system from 2 to 1, which is demonstrated by the fact that the wave functions  \pref{pxeigen} are characterized by a single momentum variable $k$. An explicit mapping from two-dimensional wave functions $\psi(x,y)$ to one-dimensional wave functions $\psi(\xi)$ can be written as
\be{transwave}
\psi(\xi)&=&\int dx dy f(\xi-x,y)\psi(x,y)\nn
\psi(x,y)&=&\int d\xi f(x-\xi,y)\psi(\xi)
\ee
with the integration kernel defined as
\be{transfunc}
f(x-\xi,y)\equiv\braket{x,y}{\xi}={1\over L}\sum_k e^{ik(x-\xi)}\psi_0(y-y_k)
\ee
In the limit $L\to\infty$ the exact expression for this function is
\be{ftrans}
f(x,y)={{\Lambda}\over{\sqrt{2\pi\sqrt{\pi}\;{\bar{l}_B}^3}}}\;\exp\left[-{1\over{2\bar l_B^2}}\Lambda x(\Lambda x+2iy)\right]
\ee
with 
\be{Lambda}
\Lambda=\left(\frac{l_B}{\bar l_B}\right)^2=\frac{\bar{\omega}_c}{\omega_c}=\sqrt{1+\frac{\omega^2}{\omega_c^2}}
\ee
We note that $\Lambda$ is effectively a scaling factor in the x-direction, and as shown by the expression \pref{Hamsp2} such a scaling is indeed needed in order for the two first terms of the Hamiltonian to have precisely the  form of the Hamiltonian for a  charged particle in the effective magnetic field $\bar B$.

The mapping between wave functions can be used to derive the corresponding mapping between operators in the 2D and 1D description. We give the expression only for the mapping of a local density interaction in 2D and refer to \cite{Horsdal07} for more details. Thus, the interaction in 2D we assume to have antisymmetric matrix elements in the coordinate representation  of the form,
\be{intpot}
 V(\v r_1,\v r_2;\v r'_1,\v r'_2)= V(\v r_1-\v r_2)(\delta(\v r_1-\v r'_1)\delta(\v r_2-\v r'_2)-\delta(\v r_1-\v r'_2)\delta(\v r_2-\v r'_1))
 \ee
 which means that it is defined by a single function of two variables, $V(x,y)\equiv V(\v r_1-\v r_2)$ with $(x,y)$ as the cartesian components of the relative position vector $\v r_1-\v r_2$ of two electrons in the plane. This operator is mapped into a {\em non-local} operator in 1D, of the form
   \be{1dinter2}
   V(\xi;\xi')=\int d^2 x \;g(\xi,\xi';x,y) V(x,y)
   \ee
where $\xi$ and $\xi'$ represent relative coordinates of the two particles in 1D. The function $g(\xi,\xi';x,y)$ is determined by the one-particle transformation function \pref{ftrans}, and has the form
\be{gtrans}
g(\xi,\xi';x,y)={\Lambda^2\over {(\sqrt{2\pi}\,\bar l_B)^3}}{\cal A} \exp\left[-{1\over{4\bar l_B^2}}\{\Lambda^2(x-\xi)^2+\Lambda^2(x-\xi')^2+2i\Lambda(\xi-\xi')y)\}\right]\nn
\ee 
with $\cal A$ denoting antisymmetrization with respect to the coordinate transformations $\xi\to -\xi$ and $\xi'\to -\xi'$.

As a special case, used in \cite{Horsdal07},  and applied also here, we consider the gaussian interaction
\be{gaussint}
V(\v r)=V_0e^{-\ga^2 \v r^2}
\ee
The corresponding 1D interaction has also a gaussian form, but depends on two rather than one length scale.  In the momentum representation the interaction has the form
\be{kspace}
V(k,k')= \bar V_0\; \cA \exp[-\frac{1}{4 a^2}(k-k')^2-b^2(k+k')^2]
\ee                   
where $k$ and $k'$ are the variables in the Fourier transform corresponding to $\xi$ and $\xi'$, while $\bar V_0$, $a$ and $b$ are constants given by
\be{rescaled}
\bar V_0&=&2\sqrt{\frac{\pi}{1+2\ga^2\bar l_B^2}}\;\frac{V_0}{\ga}\nn
a&=&\Lambda\frac{\ga}{\sqrt{\Lambda^2+2\ga^2\bar l_B^2}} \nn
b&=&\frac{\ga\bar  l_B^2}{\Lambda\sqrt{1+2\ga^2\bar  l_B^2}} 
\ee

\section{ Low energy Hamiltonian and correlation function in 1D}
In the 1D representation the second quantized Hamiltonian  has the form
\be{secondHam}
H=\sum_k({{\hbar^2}\over{2M}}k^2+\half\hbar\bar{\omega}_c)c_k^{\dag}c_k+\frac{1}{4L}\sum_{q,k_1,k_2}V(k,k')c_{k_1}^{\dag}c_{k_2}^{\dag}c_{k_2-q}c_{k_1+q}
\ee
with $k=(k_1-k_2)/2$ and  $k'=(k_1-k_2)/2+q$, with $M=m\;\bar\omega_c^2/\omega^2$  as the effective 1D mass and with $V(k,k')$ given by \pref{kspace}. The operators $c_{k}^{\dag}$ and $c_{k}$ denote in the usual way electron creation and annihilation operators. Assuming a weakly interacting system, a low-energy form of this Hamiltonian can be found.  Thus, with the assumption that excitations are energetically restricted to states close to the Fermi points $\pm k_F$, \ie to states with $k$-values restricted by $|k\pm k_F|<<k_F$,  the non-local interaction \pref{kspace} can be represented by two different local density interactions, $V_1(q)$ between particles at the same Fermi point and $V_2(q)$ between particles at opposite Fermi points. For the gaussian interaction these are given by \cite{Horsdal07}
\be{V1V2}
V_1(q)=\bar V_0\exp[-\frac{q^2}{4a^2}]\;,\quad V_2(q)=\exp[-4b^2k_F^2]\;V_1(q)
\ee
If furthermore the first term of \pref{secondHam} is linearized in $k$ around the Fermi points, the Hamiltonian can be brought into the Luttinger form 
\be{fullHam}
H=\bar v_F\hbar \sum_{\chi ,\,k}(\chi k-k_F):c_{\chi,\, k}^{\dag}\,c_{\chi,\, k}:+\frac{1}{4L} \sum_{\chi,\,q}(V_1(q)\rho_{\chi,\,q}\rho_{\chi,\,-q}+V_2(q)\rho_{\chi,\,q}\rho_{-\chi,\,-q})\nn\ee
where the $c^{\dag} c$ operator and the particle density operators $\rho_{\chi,\,q}$ are normal ordered relative to the non-interacting ground state. In this expression $\chi$ denotes the chiral quantum number which initially is defined with $\chi=+1$ for positive $k$ and $\chi=-1$ for negative $k$.  However, this relation between $k$ and $\chi$ can subsequently be relaxed, so that in \pref{fullHam} $k$ runs from $-\infty$ to $+\infty$ for both chiralities, without the low energy part of the theory being affected. Furthermore, in the linearized kinetic term the Fermi velocity has been introduced in the form
\be{Fermiv}
\bar v_F=v_F-\frac{\bar V_0 }{4\pi\hbar}(1-e^{-4b^2k_F^2})
\ee
where $v_F=k_F/M$ is the unperturbed Fermi velocity and where the interaction-dependent contribution to $\bar v_F$ arises from the modification of the background potential due to presence of the Fermi sea \cite {Horsdal07}.

The bosonization technique can be used to diagonalize the Hamiltonian \pref{fullHam} and to find the electron correlation function of the interacting ground state. As stressed in \cite{Haldane81} one needs to consider the $q=0$ component of the charge density separately and to identify the $q\neq 0$ of the charge density with the bosonic operators. In our notation the relations are
\be{crean}
a_q=\sqrt{{2\pi}\over{|q|L}}\sum_\chi\theta(\chi\,q)\rho_{\chi,q}\;,\quad a_q^{\dag}=\sqrt{{2\pi}\over{|q|L}}\sum_\chi\theta(\chi\,q)\rho_{\chi,-q}\quad (q\neq 0)
\ee
where $a_q$ and $a_q^{\dag}$ satisfy standard boson commutation relations.

Expressed in terms of the bosonic operators the Hamiltonian takes the form
\be{fullHam2}
H&=&{{\pi\hbar}\over{2L}}(v_N(N-N_0)^2+v_JJ^2)
\nn
&+&{\hbar\over 2}\sum_{q\neq 0}|q|\left[
(\bar v_F+\frac{V_{1}(q)}{4\pi\hbar})
(a_q^{\dag}a_q+a_q a_{q}^{\dag})+\frac{V_{2}(q)}{4\pi\hbar}(a_q^{\dag}a_{-q}^{\dag}+a_q a_{-q})\right] \nn
\ee
where unessential terms (constants or terms proportional to the particle number) have been subtracted. In the above expression we have included the particle number $N$ and the chiral charge $J$,
\be{NJ}
N=N_0+ \sum_{\chi }\rho_{\chi,\,0}\;,\quad J= \sum_{\chi }\chi \;\rho_{\chi,\,0}
\ee
with $N_0$ as the particle number of the ground state.
The two new velocity parameters that are introduced are
\be{velocities}
v_N&=&\bar v_F+ \frac{\bar V_0}{4\pi\hbar}(1+ e^{-4b^2k_F^2})=v_F+\frac{\bar V_0}{2\pi\hbar}e^{-4b^2k_F^2}\nn
v_J&=&\bar v_F+ \frac{\bar V_0}{4\pi\hbar}(1-e^{-4b^2k_F^2})=v_F
\ee
where only $v_N$ is modified by the interactions, as discussed in \cite{Horsdal07}.

The Hamiltonian \pref{fullHam2} is in the standard way diagonalized by a Bogoliubov transformation, which transforms the  annihilation operators as
\be{Bog}
a_q=S^{\dag}b_qS=\cosh\gz_q\; b_q+\sinh\gz_q\;b_{-q}^{\dag}
\ee
with
\be{S}
S=\exp[\half\sum_{q\neq 0} \gz_q\,(a_q^{\dag}a_{-q}^{\dag}-a_qa_{-q})]
\ee
and $\gz_q$ defined by
\be{tanh}
\tanh(2\gz_q)=-\frac{V_{2}(q)}{V_{1}(q)+4\pi\hbar \bar v_F}=-e^{-4b^2k_F^2}\frac{\bar V_0 \exp(-\frac{q^2}{4a^2})}{\bar V_0 \exp(-\frac{q^2}{4a^2})+4\pi \hbar \bar v_F}
\ee
Expressed in terms of the transformed bosonic operators the Hamiltonian is
\be{transHam}
H=
\hbar \sum_{q\neq 0}\omega_qb_q^{\dag}b_q+{{\pi\hbar}\over{2L}}(v_N(N-N_0)^2+v_JJ^2)
\ee
where again a constant has been subtracted.
The Hamiltonian has a free field form, with the boson frequency given by
\be{omega}
\omega_q=\frac{1}{4\pi\hbar}\sqrt{(V_{1}(q)+4\pi \hbar \bar v_F)^2-V_{2}(q)^2}\;|q|
\ee

 The bosonization technique makes it possible to determine the electron correlation of the 1D interacting system. By use of the mapping  between the 1D and 2D system it also determines the correlation functions of the quantum Hall state. We examine this in the following and in particular compare the asymptotic behaviour of the correlation function in 1D and 2D. We also examine the effect of the interaction on the particle density.
 
 The chiral (equal-time) electron correlation function of the 1D system is 
 \be{chicorr}
 C_{\chi}(\xi-\xi')=\bra{0}S\psi_{\chi}^{\dag}(\xi)\psi_{\chi}(\xi')S^{\dag}\ket{0}
 \ee
where $\psi_{\chi}(x)$ is the chiral electron annihilation operator, $\ket{0}$ is the ground state of the non-interacting Hamiltonian and $S$ gives the transformation between ground states of the non-interacting and the interacting Hamiltonian. 

Based on the bosonic form of the electron operators and the commutation relations between the bosonic creation and annihilation operators and $S$, an explicit expression for the correlation function in the interacting case can be found.  We refer to \cite{Haldane81} for more details (note however a difference in sign convention), and only cite the expression for the chiral correlation function,
\be{chicorr2}
 C_{\chi}(\xi-\xi')=-{{\chi}\over{2\pi i}}\;\frac{e^{-i\chi k_F(\xi-\xi')}}{\xi-\xi'+i\chi \epsilon}\,e^{-A(\xi-\xi')}
 \ee
with $\epsilon\ra0^+$ and where the function $A(\xi)$ is determined by the interaction potential as
\be{Afunc}
A(\xi)&=&4\sum_{q>0}{{2\pi}\over{Lq}}\sinh^2\gz_q\sin^2(q\,\xi/2)
\ee

The asymptotic behaviour of the correlation function, for large $\xi$, can be found by considering the derivative of the function $A(\xi)$, which in the limit $L\ra\infty$ is
\be{Ader}
\dd{A(\xi)}{\xi}=2\int_0^{\infty} dq\; \sinh^2\gz_q\sin(q\xi)
\ee
For large $\xi$ the integral collects the main contribution from small $q$, and we get the asymptotic approximation
\be{asympt}
\dd{A(\xi)}{\xi}\approx \frac{\gg-1}{\xi}
\ee
with
\be{gamma}
\gamma=\frac{\bar V_0+4\pi \hbar \bar v_F}{\sqrt{(\bar V_0+4\pi \hbar \bar v_F)^2-\exp({-8b^2k_F^2})\bar V_0^2}}
\ee
The corresponding asymptotic form of the chiral correlation function is
\be{chiasymp}
 C_{\chi}(\xi)\approx K \chi  e^{-i\chi k_F \xi}\;\xi^{-\gg}
 \ee
with $K$ as a constant.  The expression shows that the deviation from the non-interacting value $\gg=1$ is determined by the strength of the interaction between the two Fermi points, which means between the two edges of the quantum Hall channel in 2D.
 
 The full electron operator can be expressed in terms of the chiral operators  when the fourier components for non-negative $k$ values are identified with the corresponding positive chirality components and the fourier components for negative $k$  are identified with  negative chirality components
 \be{electron}
 c_k= c_{+,\,k} \quad {\rm if} \;\;k\geq 0 \nn
  c_k= c_{-,\,k} \quad {\rm if}\;\; k< 0 
\ee
 This gives the following transformation between the chiral and non-chiral electron wave functions
 \be{electron2}
 \psi(\xi)=\sum_{\chi}\int_{-{L\over 2}}^{{L\over 2}}F_{\chi}(\zeta)\;\psi_{\chi}(\xi+\zeta) d\zeta
 \ee
 with
 \be{Ffunc}
 F_{\chi}(\zeta)=\frac{\chi}{2iL} \; \frac{e^{i{{\pi}\over L}\zeta}}{\sin({{\pi}\over L}\zeta)-i\chi\epsilon}
 \approx \frac{\chi}{2\pi i} \; \frac{1}{\zeta-i\chi\epsilon}
 \ee
where the last expression corresponds to the $L\to \infty$ limit. 
 Since the $F_{\chi}$ function satisfies the following property
 \be{Ffuncprop}
 F_{\chi}(\zeta)=\int d\xi F_{\chi}(\xi+\half \zeta) F_{\chi}^*(\xi-\half \zeta)
 \ee
the  corresponding mapping between the chiral and non-chiral electron correlation functions can be expressed  as
\be{electroncorr}
C_{1D}(\xi)&=&\sum_{\chi}\int d\zeta F^*_{\chi}(\zeta)C_{\chi}(\xi+\zeta)\nn
&=&-\sum_{\chi}\frac{1}{4\pi^2}\int d\gz \;\frac{1}{\gz+i\chi\epsilon}\;\frac{e^{-i\chi k_F(\xi+\gz)}}{\xi+\gz+i\chi\epsilon'}\;e^{-A(\xi+\gz)}
\ee
To evaluate this integral we express the factor $e^{-A(\gz+\xi)}$ in terms of its Fourier components and perform the $\xi$ integral for each Fourier component by closing the integration contour either in the upper or the lower half plane, depending on the sign of the $\xi$-dependent  exponent in the integrand. The result can be written as
\be{electroncorr2}
C_{1D}(\xi)=-\sum_{\chi}\frac{\chi}{4\pi^2}(\frac{e^{-i\chi k_F \xi}}{\xi+i\chi\epsilon}-\frac{1}{\xi+i\chi\epsilon})\;e^{-\hat A(\xi)}
\ee
where the factor $e^{-\hat A(\xi)}$ is derived from $e^{-A(\xi)}$ by canceling the contributions from the high frequency Fourier components for $k<-k_F$ (or $k>k_F)$. Assuming these high frequency components to be unessential we have $\hat A(\xi)\approx A(\xi)$ and the 1D electron correlation function gets the form
\be{electroncorr3}
C_{1D}(\xi)=\sum_{\chi}C_{\chi}(\xi)-\delta(\xi)={1\over\pi}\frac{\sin k_F\xi}{\xi}e^{-A(\xi)}
\ee
in accordance with the corresponding expression in \cite{Haldane81}.

\section{The electron correlation function in 2D}
We focus now on the  problem of finding the electron correlation function for the 2D quantum Hall system. This function can be determined from the 1D electron correlation function by use of  the explicit mapping that exists between the 1D to 2D systems. 

Before proceeding we introduce two simplifications. The first one is to set $\Lambda=1$. This can either can be seen as an approximation which is valid when $\omega<<\omega_c$, or it can be viewed as a change to rescaled variables in the x-direction,  where $\Lambda$ is absorbed in the the new coordinates $x$ and $\xi$. (However, we shall use the same notations for these variables as before). Since in the following we only refer to the magnetic length of the {\em effective} magnetic field, we also introduce the simplification of referring to this length  as $l_B$ rather than $\bar l_B$. 

The correlation function of the quantum Hall state, which is translationally invariant in the x-direction, we write as
\be{2Dcorrfunc}
C_{2D}(x;y,\eta)=\mean{\psi^{\dag}(x,y+\eta /2)\psi(0,y-\eta/2)}
\ee
This is related to the 1D  correlation function by a transformation of the form
\be{2D1D}
C_{2D}(x;y,\eta)=\int d\xi \;\phi(\xi;y,\eta)\;C_{1D}(x+\xi)
\ee
where the transition function $\phi$ is determined from the $f$-functions as
\be{phifunc}
\phi(\xi;y,\eta)&=&\int dx\,f^*(x-\xi/2,y+\eta/2)\;f(x+\xi/2,y-\eta/2)\nn
&=&\frac{1}{2\pi l_B^2}\exp[-\frac{1}{4l_B^2}(\xi(\xi+4iy)+\eta^2)]
\ee
This gives the following integral expression for the correlation function
\be{2Dcorrfunc2}
C_{2D}(x;y,\eta)=\frac{1}{2\pi^2 l_B^2}\int d\xi\,\exp[-\frac{1}{4l_B^2}(\xi(\xi+4iy)+\eta^2)]
\frac{\sin [k_F(x+\xi)]}{x+\xi}e^{-A(x+\xi)}
\ee

With the interaction dependent function $A(\xi)$ given by \pref{Afunc} this integral cannot be evaluated explicitly. However, we note the presence of the gaussian function that effectively restricts the integration to an interval of length $l_B$. If the integrand is rewritten as the product of this  gaussian function and a function that varies slowly over the length scale $l_B$, the latter can be expanded about the centre of the gaussian to give a rapidly converging series where the lowest order terms can be evaluated.
Since the function $\sin [k_F(x+\xi)]$ is rapidly oscillating, it is useful to rewrite the integrand in terms of the two chiral contributions,
\be{2Dcorrfunc3}
C_{2D}(x;y,\eta)=-\frac{1}{4\pi^2 i l_B^2}\sum_{\chi} \chi \{\exp[-\frac{1}{4l_B^2} \eta^2-i\chi k_F x]\nn
\times \int d\xi\,\exp[-\frac{1}{4l_B^2}(\xi^2+4i\xi y+4i\chi k_Fl_B^2\xi)]
\frac{e^{-A(x+\xi)}}{x+\xi\pm i\epsilon}\}
\ee
where a shift $\pm i \epsilon$ ($\epsilon\to 0^+$) has been introduced for the location of the poles relative to the real axis. Note that the sign of this shift can be chosen freely, but has to be the same for both chiralities in the sum.

We introduce a new, $\chi$-dependent, variable in the y-direction, $y_{\chi}= y+\chi k_Fl_B^2$. For $\chi=+1$ it measures $y$ relative to the edge located at $-k_Fl_B^2$ and for $\chi=-1$ it measures $y$ relative to the other edge at $+k_Fl_B^2$. We also introduce a new, $\chi$-dependent,  integration variable $\xi'= \xi+2iy_{\chi}$. This gives the following expression for the correlation function,
\be{2Dcorrfunc4}
C_{2D}(x;y,\eta)=-\frac{1}{4\pi^2 i l_B^2}\sum_{\chi}\chi\{\exp[-\frac{1}{4l_B^2} (\eta^2
+4y_{\chi}^2)-i\chi k_F x]\nn
\times \int_{C_{\chi}} d\xi'\,\exp[-\frac{1}{4l_B^2}\xi'^2]
\frac{e^{-A(x+\xi'-2i y_{\chi})}}{x+\xi'-2i y_{\chi}\pm i\epsilon}\}
\ee
with $C_{\chi}$ indicating an integration path shifted by $2i y_{\chi}$ relative to the real axis. This integration path can be shifted back to the real axis, but  due to the crossing of poles through the integration path during the shift, there will be compensating pole contributions. We therefore write the result as
\be{2Dcorrfunc5}
C_{2D}(x;y,\eta)=C_{int}(x;y,\eta)+C_{pole}(x;y,\eta)
\ee
with the integral contribution given by integrals along the real axis,
\be{corrint}
C_{int}(x;y,\eta)=-\frac{1}{4\pi^2 i l_B^2}\sum_{\chi}\chi\{\exp[-\frac{1}{4l_B^2} (\eta^2
+4y_{\chi}^2)-i\chi k_F x]\nn
\times \int d\xi\,\exp[-\frac{1}{4l_B^2}\xi^2]
\;\frac{e^{-A(x+\xi-2i y_{\chi})}}{x+\xi-2i y_{\chi}}\}
\ee
while the pole contribution is found to be
\be{corrint2}
C_{pole}(x;y,\eta)=\left\{ {\matrix{{0 
\quad\quad\quad\quad\quad\quad\quad\quad\quad\quad
\quad\quad\quad\quad\quad\quad\quad   
y>k_Fl_B^2 }
\cr{\frac{1}{2\pi  l_B^2}\exp[-\frac{1}{4l_B^2} (x^2+\eta^2-4ixy)]
\quad -k_Fl_B^2<y<k_Fl_B^2}
\cr {0\quad\quad\quad\quad\quad\quad\quad\quad\quad\quad
\quad\quad\quad\quad\quad\quad\quad
y<-k_Fl_B^2} }} \right.
\ee

\begin{figure}[h]
\label{veier}
\begin{center}
\includegraphics[height=3.5cm]{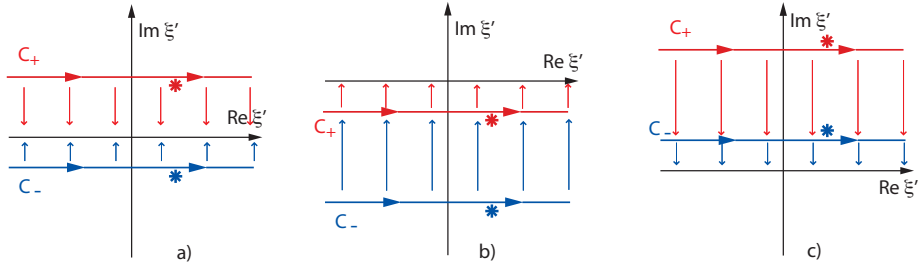}
\end{center}
\caption{\small The integration paths for three different cases, a) $y$ lies between the edge values $\pm k_Fl_B^2$, b) $y$ is smaller than $- k_Fl_B^2$ and c)  $y$ is greater than  $+ k_Fl_B^2$. $C_+$ denotes the integration path for the positive chirality term and $C_-$ for the negative chirality term, and the location of the poles of the integrands is indicated  for each of these by the nearby star. In case a) the integration paths, when shifted to the real axis (vertical arrows), have to be compensated for by a pole contribution from the pole close to $C_+$, while in the cases b) and c) the paths can be shifted to the real axis without passing the the associated poles. }
\end{figure}

We note that there is no pole contribution for $y>k_Fl_B^2$ and for $y<-k_Fl_B^2$. This is most easily demonstrated in \pref{2Dcorrfunc4} by choosing the shift $-i\epsilon$ in \pref{2Dcorrfunc4}  for $y>k_Fl_B^2$ and $+i\epsilon$ for $y<-k_Fl_B^2$. The integration paths then can be shifted to the real axis without crossing the poles. In the case $ -k_Fl_B^2<y<k_Fl_B^2$ a crossing of one of the poles cannot be avoided, and that gives rise to the pole contribution (see Fig.~1). 

The pole contribution $C_{pole}$, which is present only for values of $y$ within the two edges, can in fact be interpreted as the bulk contribution to the correlation function, since it is independent of the location of the edges. The edge contributions are then given by $C_{int}$. As shown by \pref{corrint2} the bulk contribution is insensitive to the electron interaction.

The discrete change in $C_{pole}$ at $y=\pm k_F l_B^2$ may suggest that the full correlation function $C_{2D}$ also changes discontinuously at the edges. However, that is not a correct conclusion, since the discontinuity of $C_{pole}$ only compensates for the discontinuous change in $C_{int}$ when one of the poles passes the integration path when $y$ is changed. Thus, $C_{2D}(x;y,\eta)$ is everywhere a continuous function of $y$.

We consider next the contribution $C_{int}$ and note that it essentially consists of two independent contributions, corresponding to the two chiralities $\chi=\pm 1$. Thus, each of these contributions is exponentially damped when $y$ moves away from the corresponding edges at $y=\mp k_F l_B^2$, and with $l_B$ as the damping length there is little overlap between the two contributions when the separation between the two edges, $W=2k_Fl_B^2$,  is much larger than $l_B$. In a similar way the integration variable is effectively limited to $\xi \lesssim 2l_B$ by the gaussian factor, so that both $y_{\chi}$ and $\xi$ are restricted to be of order $l_B$.

The function $A(x)$, which depends on the particle interaction, is a slowly varying function of $x$, with the interaction length $1/\ga$ as the typical scale on which the function changes. Since this is assumed to be much longer than $l_B$, an expansion of $A(x+\xi-2iy_{\chi})$ in powers of $\xi-2iy_{\chi}$ should give rise to a rapidly converging series. Here we restrict the expansion to second order,
\be{Aexpand}
A(x+\xi-2iy_{\chi})= A(x)+\ga(\xi-2iy_{\chi})B(x)+\half \ga^2(\xi-2iy_{\chi})^2C(x)+...
\ee
where the higher derivatives of $A(x)$ are expressed through the dimensionless functions $B(x), C(x),...$. This is essentially an expansion in powers of the small quantity $\ga l_B$.

When the expansion is restricted to second order, the exponent of the integrand in \pref{corrint} is quadratic in the the integration variable $\xi$, like in the non-interacting case, and the the effect of the interaction can then be absorbed in the $x$ and $y$ dependent coefficients,
\be{corrint3}
C_{int}(x;y,\eta)=-\frac{1}{4\pi^2 i l_B^2}\sum_{\chi}\chi\,\{\exp[-\frac{1}{4l_B^2} (\eta^2
+4y_{\chi}^2)-i\chi k_F x-A(x)]\nn
\times \int d\xi\,\frac{\exp[-\frac{1}{4l_B^2}\xi^2-\ga(\xi-2iy_{\chi})B(x)-\half\ga^2(\xi-2iy_{\chi})^2C(x)]}{x+\xi-2i y_{\chi}}
\ee
By introducing a shift in the  variable $\xi$ the integral can be written in the form
\be{corrint4}
C_{int}(x;y,\eta)=-\frac{1}{4\pi^2 i l_B^2}\sum_{\chi}\chi\,e^{\Phi}
\int_{-\infty}^{\infty}dt\,\frac{e^{-t^2}}{t+z}
\ee
with coefficients
\be{coeff}
\Phi &=& -\frac{1}{4l_B^2} (\eta^2
+4y_{\chi}^2)-A(x)+2 \ga^2y_{\chi}^2C(x)+\ga^2l_B^2B(x)^2+i\chi k_F x+2i\ga y_{\chi} B(x)\nn
z&=&\frac{x}{2l_B}(1+\ga^2l_B^2C(x))-i\frac{y_{\chi}}{l_B}(1-\ga^2l_B^2C(x))-l_B\ga B(x)
\ee
where terms of higher than second order in $\ga l_B$ are neglected.

The integral in \pref{corrint4} can be expressed in terms of the error function in the following way,
\be{errorfunc}
\int dt\,\frac{e^{-t^2}}{t+z}=-i\pi e^{-z^2}({\rm erf}(iz)+{\rm sgn}({\rm Im}(z))
\ee
where the term proportional to ${\rm sgn}({\rm Im}(z))$ has a discontinuous $y$-dependence that exactly  matches the omitted term $C_{pole}$. To demonstrate this, we note that the corresponding contribution to $C_{int}$ can be written as
\be{discont}
\frac{1}{4\pi l_B^2}\sum_{\chi}\chi e^{\Phi-z^2}{\rm sgn}({\rm Im}(z))&=&-\frac{1}{4\pi l_B^2}\sum_{\chi}\chi\exp[-\frac{1}{4l_B^2} (x^2+\eta^2-4ixy)]{\rm sgn}(y_{\chi})\nn
&=& -C_{pole}(x;y,\eta)
\ee
To obtain the correct exponent on the right hand side of this equation one has to evaluate $\Phi-z^2$ to second order in $\ga l_B$, consistent with the approximation we apply, and to note that $x$ should be treated as of order $l_B$, due to the exponential damping factor. 

For the contribution to $C_{int}$ that is proportional to ${\rm erf}(iz)$ the situation is different.  There is no exponential damping of this term with increasing $x$ since the factor $\exp(-x^2/l_B^2)$ is compensated by a corresponding exponential increase of the error function. The overall dependence of this term with $x$ is therefore a non-exponential decrease.

For the full electron correlation function we now get the following expression, 
\be{fullcorr}
C_{2D}(x;y,\eta)=\frac{1}{4\pi l_B^2}\{{\rm erf}[i\;z_+(x,y)]-{\rm erf}[i\;z_-(x,y)]\}e^{-\Delta(x;y,\eta)}
\ee
with 
\be{zpm}
z_{\pm}(x,y)=
(1+\ga^2l_B^2C(x))\frac{x}{2l_B}-l_B\ga B(x)-i(1-\ga^2l_B^2C(x))\frac{y\pm k_Fl_B^2}{l_B}
\ee
and
\be{delta}
\Delta(x;y,\eta)=(1+2\ga^2l_B^2C(x))\frac{x^2}{4l_B^2}+A(x)-\ga B(x) x+\frac{\eta^2}{4l_B^2}-\frac{i}{l_B^2}xy
\ee
This final expression for the correlation function includes effects of the electron interaction up to second order in $\ga l_B$.

\begin{figure}[h]
\label{corrfig2}
\begin{center}
\includegraphics[height=8cm]{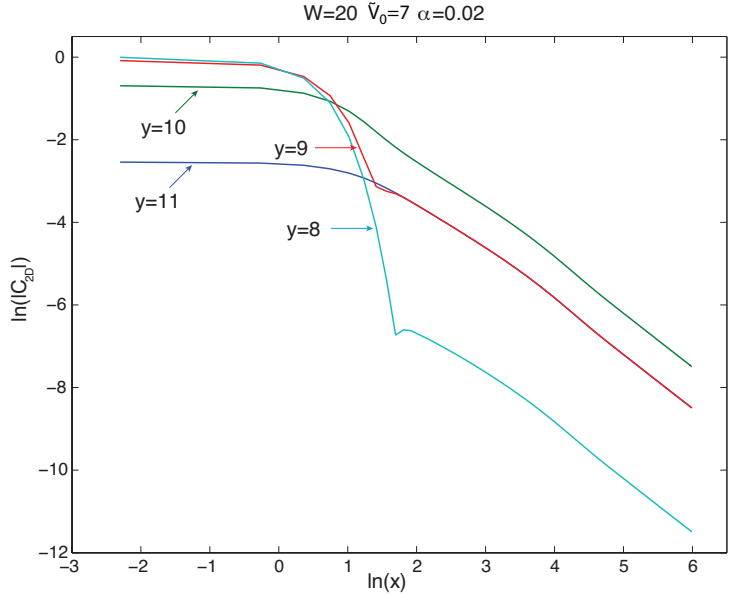}
\end{center}
\caption{\small The absolute value of the electron correlation function $C_{2D}(x,y,\eta)$ as function of the relative coordinate $x$ for different values of of the $y$ coordinate. In all cases the $y$-coordinate is chosen to be the same for the two points of the correlation function ($\eta=0$). The value $y=10$ corresponds to the edge value $k_F l_B^2$ with the chosen units.  }
\end{figure}

In Fig.~2 we have illustrated the form of the electron correlation by plotting the absolute value of $C_{2D}$ as a function of $x$ in a log-log plot for a particular choice of the interaction strength and interaction length. The function is shown for different $y$-values, corresponding to points outside the edge, at the edge and inside the edge. For $y<k_F l_B^2$ (which is $10$ in the chosen units) the curves show a clear distinction between the short range  bulk contribution, with a rapid fall-off with $x$, and the long range edge contribution with a slower, almost linear change with $x$. 

In Fig.~2, as well as in Figs.~3 and 4, the width W is measured in units of magnetic length, $\tilde V_0$ gives the value of the dimensionless interaction strength $\bar V_0/4\pi\hbar \bar v_F$ and the inverse interaction length $\alpha$ is measured in units of inverse magnetic length. The correlation function $C_{2D}$ is measured in units of the bulk density $1/2\pi l_B^2$ and the coordinates $x$ and $y$ are measured in units of the magnetic length.
\section{Electron density and the asymptotic form of the correlation function}
The electron density is determined by the correlation function \pref{fullcorr} for $x=\eta=0$. Since the function $A(x)$ is proportional to $x^2$ for small $x$, we have $A(0)=B(0)=0$, and the electron density is
\be{density}
\rho_{2D}(y)=\frac{1}{4\pi l_B^2}\{{\rm erf}[(1-\ga^2l_B^2C)\frac{y+k_Fl_B^2}{l_B}]-{\rm erf}[(1-\ga^2l_B^2C)\frac{y- k_Fl_B^2}{l_B}]\}
\ee
with $C=C(0)$. This constant is determined by the second derivative of $A(x)$, and within the approximation applied here the explicit expression is
\be{C0}
C=2\int_0^{\infty} dz \left(\frac{\bar V_0e^{-z}+4\pi\hbar \bar v_F}{\sqrt{(\bar V_0e^{-z}+4\pi\hbar \bar v_F)^2-e^{-2\ga^2 W^2}\bar V_0^2e^{-2z}}}-1\right)
\ee 
Thus, the electron interaction gives rise to a small change in the density, which is of second order in $\ga l_B$. In the non-interacting case the density profile has the well-known form with constant density, corresponding to a full lowest Landau level, for $y$ values between the edges at $y=\pm k_F l_B^2$. 
We see from the expression \pref{density} that the effect of the interaction is to widen the edge by introducing a slightly larger length scale $l_B\to l_B/((1-\ga^2l_B^2C)$, while the bulk density is left unchanged. For small $\ga$ this length increases with $\ga$, but for sufficiently large $\ga$ it is exponentially damped due to the presence of the factor $\exp(-2\ga^2 W^2)$ in the expression for $C$. 

\begin{figure}[h]
\label{densityfig}
\begin{center}
\includegraphics[height=5.5cm]{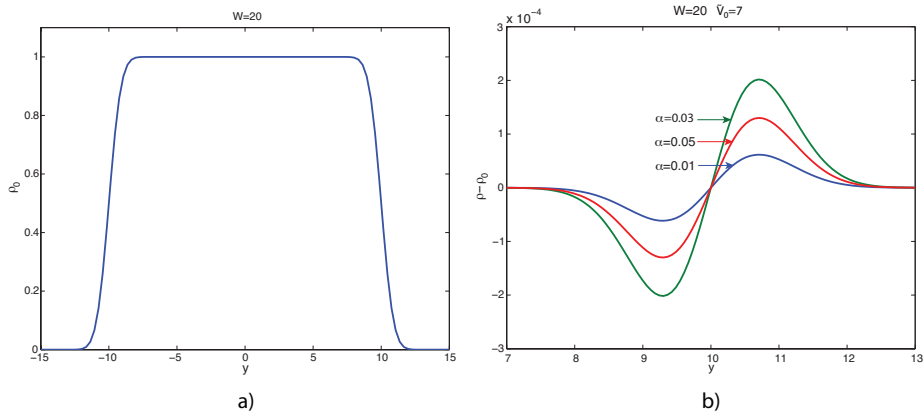}
\end{center}
\caption{\small The electron density as a function of the transverse coordinate $y$. Plot a) shows the density profile across the 2D system, with an almost constant density between the two edges. Plot b) shows the effect of the interaction, in the form of the difference between the density of the interacting and the non-interacting system for values of $y$ close to the edge (at $y=10$). The three curves correspond to three different values of the interaction range $\alpha$, with fixed value of the interaction strength $\tilde V_0$. The curve with largest deviation from the non-interacting case corresponds to $\alpha W = 0.6$.  }
\end{figure}

The electron density is shown in Fig.~3 for a fixed interaction strength and variable interaction lengths. Fig.~3a shows the standard picture of a quantum Hall fluid, with a constant electron density in the interior of the 2D channel, corresponding to integer filling, and with an abrupt fall in the density  to 0 at the edges. The width of the edge is essentially equal to the magnetic length $l_B$. For the chosen value of the interaction strength the effect of the interaction is not detectable in Fg.~3a, but is demonstrated in Fig.~3b where a magnified picture of the density is shown for y-values close to the edge. The graph shows deviations from the density of the non-interacting system for three different values of the interaction length. The maximal effect corresponds to an interaction length that approximately matches the width of the channel.

We next consider the asymptotic form of the correlation function given by \pref{fullcorr}. We make use of the leading terms of the asymptotic expansion for the error function \cite{Gradshteyn80}
\be{asymptot}
e^{-z^2}{\rm erf}(iz)=e^{-z^2} + \frac{i}{\sqrt{\pi}z}+\cO(\frac{1}{z^3})
\ee
and the asymptotic expressions
\be{asymptot2}
A(x)&{\approx}& (\gamma-1)\ln(x/x_0)\nn
B(x)&{\approx}& (\gamma-1)\frac{1}{\ga x}\nn
C(x)&{\approx}& -(\gamma-1)\frac{1}{\ga^2 x^2}
\ee
with $x_0$ as a constant. This gives 
\be{corrasymp}
C_{2D}(x;y,\eta)&\approx& \frac{i}{4\pi \sqrt\pi l_B^2}(\frac{e^{-\Delta+z_+^2}}{z_+}-\frac{e^{-\Delta+z_-^2}}{z_-})\nn
&\approx &\frac{i }{2\pi\sqrt{\pi}l_B x_0}(x/x_0)^{-\gamma}\exp(-\frac{\eta^2}{4l_B^2})\nn
&&\times(\exp(-\frac{((y+k_Fl_B^2)^2}{l_B^2}-ik_F x)
-\exp(-\frac{(y-k_Fl_B^2)^2}{l_B^2}+ik_F x))\nn
\ee
Near one of the edges, \eg $y \approx k_F l_B^2$, this gives the asymptotic expression
\be{corrasymp2}
|C_{2D}(x;y,\eta)|\approx K' x^{-\gamma}\exp(-\frac{((y-k_Fl_B^2)^2}{l_B^2})
\exp(-\frac{\eta^2}{4l_B^2}
)
\ee
with $K'$ as a new constant. The expression shows that the asymptotic $x$-dependence of the 2D electron correlation function is determined by the same exponent $\gamma$ as the  $\xi$-dependence of the 1D (chiral) correlation function, \pref{chiasymp}. It also shows that the variation of the function in the y-direction has a gaussian form, with the function peaking at the edge of the system. 

\begin{figure}[h]
\label{corrfig4}
\begin{center}
\includegraphics[height=9cm]{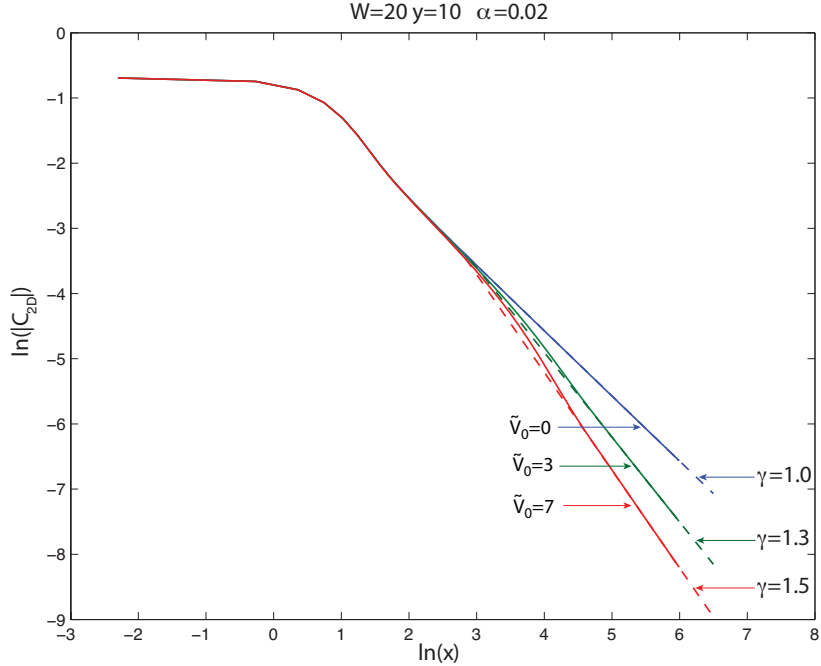}
\end{center}
\caption{\small The absolute value of the correlation function $C_{2D}$ as a function of the coordinate $x$ for fixed $y=k_F l_B$ (equal to 10 in the chosen units) and with $\eta =0$. The interaction range is fixed, with  $\alpha l_B=0.02$, and the three curves correspond to three values of the interaction strength $\tilde V_0$, where one of them is the non-interacting case. The dashed curves are straight lines with slopes determined by the values of the asymptotic parameter $\gg$ for the three cases. The plotted curves fit well the asymptotic lines for large $x$, but one may note that there are deviations for $x$ values that are smaller than about 50 magnetic lengths, which is comparable to the range of the interaction potential.}
\end{figure}

In Fig.~4 we illustrate the asymptotic form of the correlation function and the transition from the small $x$ behaviour, with the three curves showing the absolute value of the correlation function as a function of $x$ for different values of the interaction strength. The constant values of the slope that are reached for large $x$ are consistent with the corresponding values of the asymptotic parameter $\gg$ in the three cases. However, we note that in the interacting cases a fairly large value of $x$ is needed before the asymptotic, linear behaviour is reproduced.

\section{Conclusions}

We have examined in this paper  the electron correlation function for a 2D quantum Hall system, which has the form of a linear channel of electrons in a strong perpendicular magnetic field. A simple model, which was introduced in Part I \cite{Horsdal07}, has been used, where the electrons are confined by a harmonic oscillator potential and are subject to a gaussian two-particle interaction. An explicit expression for the 2D correlation function has been derived by use of a mapping of this system to a 1D Luttinger liquid, a mapping that was discussed in detail in  \cite{Horsdal07}. 

We have made use of the low-energy form of the theory in 1D, where the electron correlation function can be determined by use of the bosonization technique, and have mapped this function back to 2D. In this way the expression for the electron correlation function of the quantum Hall system has been found, which is valid within the low-energy approximation. The expression found in this way shows a natural distinction between a bulk contribution and an edge contribution to the correlation function.

The method used here (and in Part I) is based on determining  how the interactions in 2D modify  the Luttinger parameters in 1D, and how this in turn gives rise to modifications in the description of  the 2D system. We find that the interactions only influence the edges of the system, since only the edge contributions to the correlation function are modified by the interactions. Thus, even if the particle interaction is long range and acts between the two edges of the system, there is only a negligible change in in the electron density and correlation function in the interior of the quantum Hall channel. For the electron density even at the edge the effect is very small, corresponding to a small widening of the edge. For the electron correlation function the main effect of the interactions is to change its asymptotic behaviour. Thus, the fall-off with the distance $x$ between the the two points of the correlation function is found to be determined by the same parameter value $\gg$ that determines the asymptotic behaviour of the 1D correlation function. However, for the specific values of the interaction strength and range that we have studied, the asymptotic form of the correlation function is reached only after a fairly long distance $x$, a distance that is comparable to the range of the interaction potential.\\

\noindent
{\bf Acknowledgement}\\
This work has financial support from the Research Council of Norway  from NordForsk.


\end{document}